\documentclass
[preprint,showpacs,byrevtex,10pt,twocolumn,tightenlines,prl,letterpaper]{revtex4}%
\usepackage{amsfonts}
\usepackage{amsmath}
\usepackage{amssymb}
\usepackage[dvips]{graphicx}
\usepackage{hyphenat}%
\providecommand{\U}[1]{\protect\rule{.1in}{.1in}}

\begin{document}
\preprint{ }
\title{Screening the Coulomb interaction and thermalization of Anderson insulators}
\author{Z. Ovadyahu}
\affiliation{Racah Institute of Physics, The Hebrew University, Jerusalem 91904, Israel }

\pacs{72.15.Rn 61.43.-j}

\begin{abstract}
Long range interactions are relevant for a wide range of phenomena in physics
where they often present a challenge to theory. In condensed matter, the
interplay of Coulomb interaction and disorder remains largely an unsolved
problem. In two dimensional films the long-range part of the Coulomb
interaction may be screened by a nearby metallic overlay. This technique is
employed in this work to present experimental evidence for its effectiveness
in limiting the spatial range of the Coulomb interaction. We use this approach
to study the effects of the long-range Coulomb interaction on the
out-of-equilibrium dynamics of electron-glasses using amorphous indium-oxide
films. The results demonstrate that electronic relaxation times, extending
over thousands of seconds, do not hinge on the long-range Coulomb interaction
nor on the presence of a real gap in the density of states. Rather, they
emphasize the dominant role played by disorder in controlling the slow
thermalization processes of Anderson insulators taken far from equilibrium.

\end{abstract}
\maketitle

\section{Introduction}

The interplay between disorder and Coulomb interaction has been a challenging
problem in condensed matter physics. Effects associated with disorder while
neglecting interaction may still be a difficult problem to solve. Such
theories however are rarely applicable for experiments as disorder and
interactions appear to be connected; increasing one usually increases the
other. Few comprehensive studies treating disorder and interactions were made,
usually when both are fairly weak or when the spatial range of the interaction
is limited. In the strong-disorder regime however, neglecting the long-range
part of the interaction is difficult to justify, thus further compounding a
difficult problem. This is true in particular for the Anderson localization
case where the question of Coulomb interaction originated decades ago
\cite{1,2} is still unsolved despite extensive efforts. Some progress has been
made on this many-body problem for short-range interaction \cite{3,4,5} but
effects of the long-range component are yet largely unresolved.

An intriguing result of the disorder-interaction competition is the appearance
of a non-ergodic phase exhibiting glassy features. These involve slow
conductance-relaxations of Anderson insulators taken far from the equilibrium
and a variety of memory effects \cite{6,7}. Relaxation times that extend over
thousands of seconds are observable at temperatures where the hopping-length,
which is the effective screening-length in the insulating regime, is of the
order of 20nm.

Theoretical models that qualitatively account for these effects are based on
the opening up of a soft-gap \cite{8,9,10,11,12,13,14,15} in the system
density of states (DOS). This, so called Coulomb-gap \cite{16}, is reflected
in the conductance G versus gate-voltage V$_{\text{g}}$ as a cusp-like minimum
centered at the point where the system was allowed to relax (the `memory-dip')
\cite{7}. To be observable in G(V$_{\text{g}}$) scans, $\partial$V$_{\text{g}%
}$/$\partial$t must be fast enough relative to the relaxation-rate of the
electronic system \cite{17}. For technical reasons this condition limits the
choice of systems to Anderson-insulators with relatively high
carrier-concentration \textit{N} where both disorder and interactions are
strong \cite{17}. The importance of strong disorder and interaction is
attested by the seven different systems that exhibit these nonequilibrium
effects all sharing the feature of high carrier-concentration \textit{N}%
$\gtrsim$10$^{\text{19}}$cm$^{\text{-3}}$. It is yet not clear however what
role is played by the long-range Coulomb interaction in these phenomena.

In this work we attempt to find answers to this and related questions by using
a metallic ground-plane in proximity to the sample to modify the long-range
Coulomb interaction in a controlled way. Using samples configured for
field-effect measurements, and furnished with a nearby screening-plane, yield
results consistent with the anticipated \cite{18} outcome for a modified
Coulomb-gap. The dynamics of these systems, on the other hand, does not show
significant difference relative to the reference samples. It seems therefore
that, in addition to strong enough quenched disorder, short and medium-range
interactions may be sufficient to account for the long relaxation times
observed in the experiments. In particular, the results demonstrate that
relaxation times extending over hours are sustainable in interacting Anderson
insulators even while having a finite density-of-states at the chemical potential.

To optimize the effect of screening by a nearby metal, the system chosen for
the study had rather low carrier-concentration. This also resulted in systems
with short relaxation-times. We took advantage of the latter to systematically
study the deviation from the logarithmic relaxation-law to elucidate the
relative importance of disorder and interaction to the slow dynamics of the
glassy phase.

\section{Experimental}

\subsection{Sample preparation and measurement techniques}

Samples used in this study were 200\AA \ thick films of In$_{\text{x}}$O.
These were made by e-gun evaporation of 99.999\% pure In$_{\text{2}}%
$O$_{\text{3}}$ onto room-temperature Si wafers in a partial pressure of
1.5x10$^{\text{-4}}$mBar of O$_{\text{2}}$ and a rate of 0.5$\pm$0.1\AA /s.
The Si wafers (boron-doped with bulk resistivity $\rho\leq$2x10$^{\text{-3}%
}\Omega$cm) were employed as the gate-electrode in\ the field-effect and
gate-excitation experiments. The samples were deposited on a SiO$_{\text{2}}$
layer (2$\mu$m thick) that was thermally-grown on these wafers and acted as
the spacer between the sample and the conducting Si:B substrate.

The as-deposited films had sheet-resistance R$_{\square}$%
$>$%
G$\Omega$ at room-temperature. They were then thermally-treated. This was done
by stages; the samples were held at a constant temperature starting from
$\approx$340K for 20-30 hours then the temperature was raised by 5-10K for the
next stage. This was repeated until the desired R$_{\square}$ was attained
(see \cite{19} for fuller details of the thermal-annealing and structure
analysis). This process yielded samples with R$_{\square}$=18-45k$\Omega$ that
at T$\approx$4K spanned the range of 100k$\Omega$ to 40M$\Omega$. The
carrier-concentration \textit{N} of these samples, measured by the Hall-Effect
at room-temperatures, was in the range \textit{N}=8.7x10$^{\text{18}}%
$cm$^{\text{-3}}$ to 2x10$^{\text{19}}$cm$^{\text{-3}}$.

The main focus in this work was a study of the effects produced by screening
the long range part of the Coulomb interaction on the nonequilibrium transport
of Anderson insulators. The experimental methodology we employed is a
comparing simultaneously deposited samples, placing a metallic-plane in close
proximity to just one of them. Figure 1 illustrates the geometry of the pair
of samples, labeled "screened" and "reference", each configured for
field-effect measurements.
\begin{figure}[ptb]%
\centering
\includegraphics[
height=1.4615in,
width=3.039in
]%
{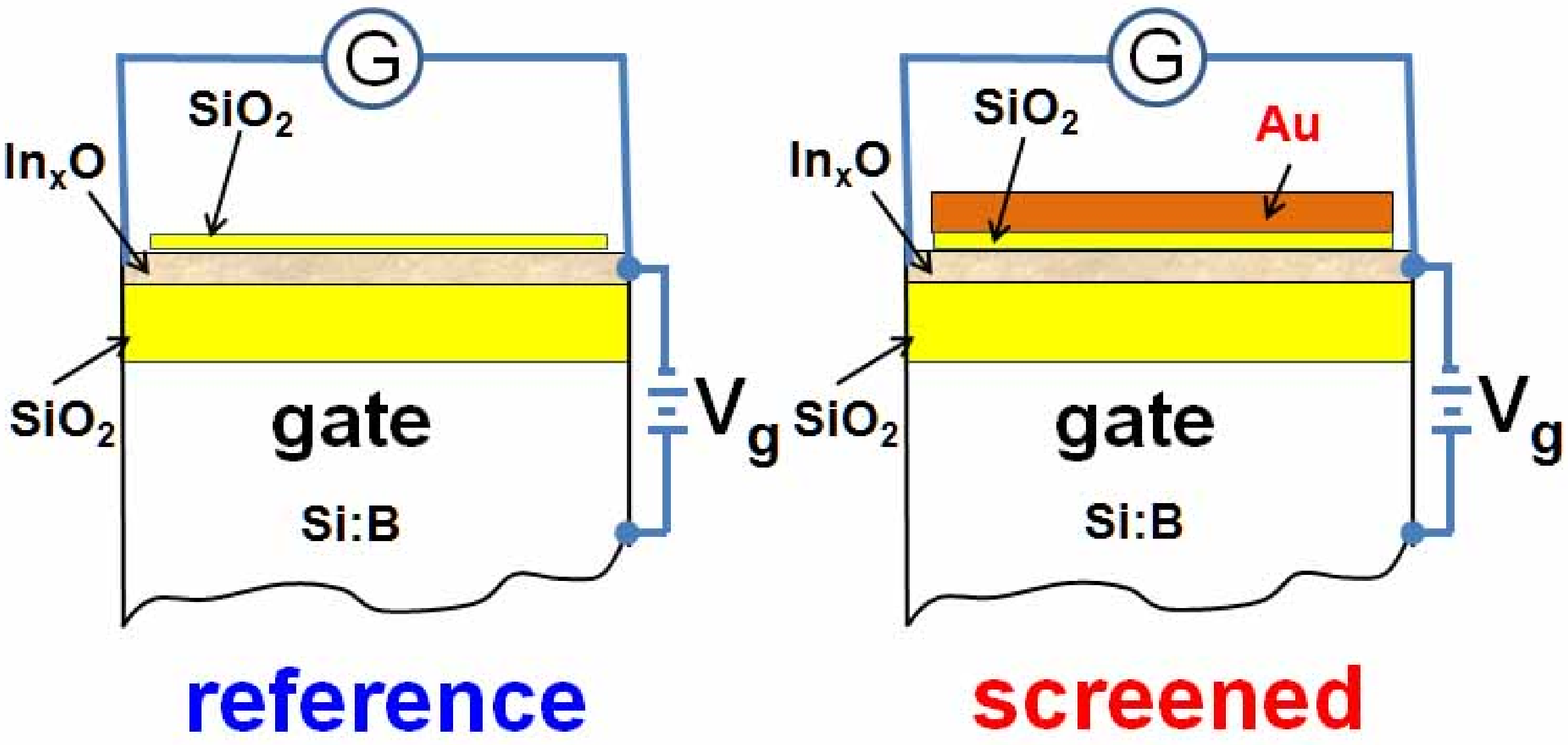}%
\caption{Schematic description of the reference and screened samples
configured for field-effect measurements.}%
\end{figure}

The distance between the screened sample and the screening-layer (a $\approx
$200\AA \ gold film) is determined by the thickness d of the SiO$_{\text{2}}$
layer. This spacer, 7-11nm thick, was e-gun deposited on both the screened and
reference sample simultaneously using pure quartz as the source. The Anderson
insulator that was chosen for these experiments was the version of
In$_{\text{x}}$O with low carrier-concentration (\textit{N}$\leq
$2x10$^{\text{19}}$cm$^{\text{-3}}$). This version has several attractive
features for these experiments: In the first place, the relatively large
inter-carrier distance \textit{N}$^{\text{-1/3}}\simeq$5nm, allows the spacer
d to be thick enough to minimize pinholes while d\textit{N}$^{\text{1/3}}$ may
still be small enough for effective screening. Secondly, the electron-glass
dynamics becomes faster as their carrier-concentration falls below
\textit{N}$\lesssim$4x10$^{\text{19}}$cm$^{\text{-3}}$ while, all other things
being equal, the relative value of the excess-conductance in the excited state
$\Delta$G/G is more conspicuous than in samples with \textit{N}$>$%
4x10$^{\text{19}}$cm$^{\text{-3}}$. These expectations were borne out in our
experiments which made it possible to quantify the system dynamics as it
approaches the quantum phase transition.

Conductivity of the samples was measured using a two-terminal ac technique
employing a 1211-ITHACO current preamplifier and a PAR-124A lock-in amplifier.
Measurements were performed with the samples immersed in liquid helium at
T$\approx$4.1K held by a 100 liters storage-dewar. This allowed up to two
months measurements on a given sample while keeping it cold. These conditions
are essential for the measurements described below where extended times of
relaxation processes are required at a constant temperature, especially when
running multiple excitation-relaxation experiment on the same sample.

The gate-sample voltage (referred to as V$_{\text{g}}$ in this work) in the
field-effect measurements was controlled by the potential difference across a
10$\mu$F capacitor charged with a constant current fed by the Keithley K220.
The rate of change of V$_{\text{g}}$ is determined by the value of this
current. The range of V$_{\text{g}}$ used in this study reached in some cases
$\pm$50V which is equivalent to the $\pm$12V used in previous studies where
the gate-sample separation was 0.5$\mu$m as compared with the 2$\mu$m
SiO$_{\text{2}}$ spacer used here.

The ac voltage bias in conductivity measurements was small enough to ensure
near-ohmic conditions. The voltage used in the relaxation experiments was
checked to be in the linear response regime by plotting the current-voltage
characteristics of each sample.

\section{Results and discussion}

\subsection{Modifying the memory-dip by a screening-plane}

The idea behind the use of the elaborate construction described in Fig.1 was
to find out the effect of eliminating (or at least weakening) the long-range
part of the Coulomb interaction. This relies on comparing results of identical
measurements on the screened and reference samples. For that to be a tenable
procedure, one has to ascertain that the two samples differ \textit{only} by
the image-charges created in the nearby gold layer. This is not a trivial
undertaking to secure as the act of depositing the gold layer may
inadvertently break the symmetry between the reference and screened samples;
For example, the heat produced during deposition of the gold\ layer will
unavoidably cause some annealing in the screened sample. A different disorder
in the screened sample may also arise from the strain related to mismatch in
mechanical properties of the Au/SiO$_{\text{2}}$ interface. In principle, a
difference in disorder between the screened-reference samples can be
compensated by a judicious thermal annealing of the samples to make their
room-temperature resistivity close to one another. However, being Anderson
insulators, a few percent difference in room-temperature resistance may
translate to orders of magnitude disparity at liquid helium temperatures.%
\begin{figure}[ptb]%
\centering
\includegraphics[
height=4.3059in,
width=3.0381in
]%
{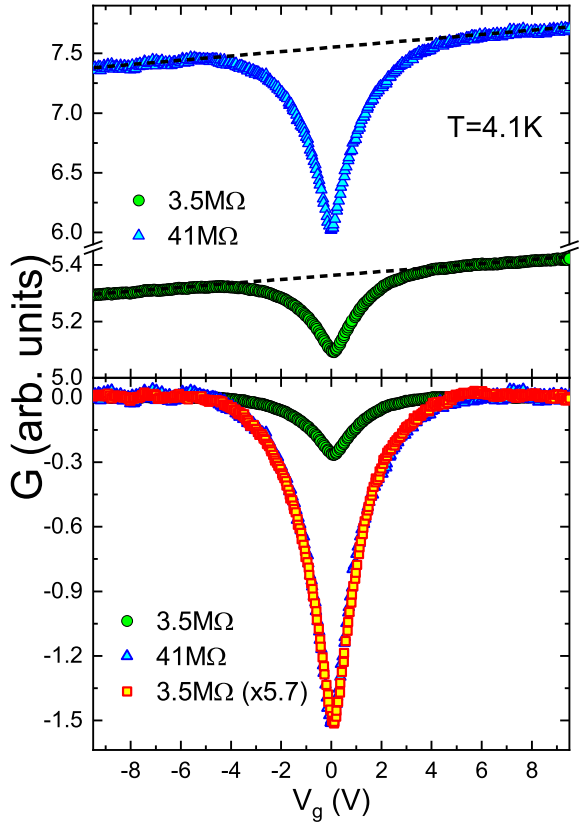}%
\caption{Field-effect plots G(V$_{\text{g}}$) for two In$_{\text{x}}$O samples
prepared from the same evaporation batch (\textit{N}=8.7x10$^{\text{19}}%
$cm$^{\text{-3}}$) but subjected to different thermal-annealing (top plate).
The dashed lines depict the thermodynamic component of the respective
G(V$_{\text{g}}$). The bottom plate shows the MD's of these samples (after
subtracting their respective thermodynamic component), and demonstrate that
their magnitude may be made to scale just by a multiplicative constant. }%
\end{figure}

Fortunately, the feature that is targeted for investigation here is not
susceptible to these artifacts; The \textit{shape} of the memory-dip which
reflects the underlying Coulomb-gap is a robust feature. At a given
temperature, the MD shape is \textit{independent} of\ the sample disorder, the
sweep-rate, time since cooldown, magnetic-fields etc., it depends
\textit{only} on the carrier-concentration which is set by the In/O ratio as
demonstrated in \cite{7}. To illustrate, figure 2 shows the dependence of the
conductance G on gate-voltage V$_{\text{g}}$ for two of the studied
`reference' samples. These share the same composition but were subjected to
different degree of annealing and thus exhibit different sheet-resistance (and
thus disorder).%
\begin{figure}[ptb]%
\centering
\includegraphics[
height=4.248in,
width=3.039in
]%
{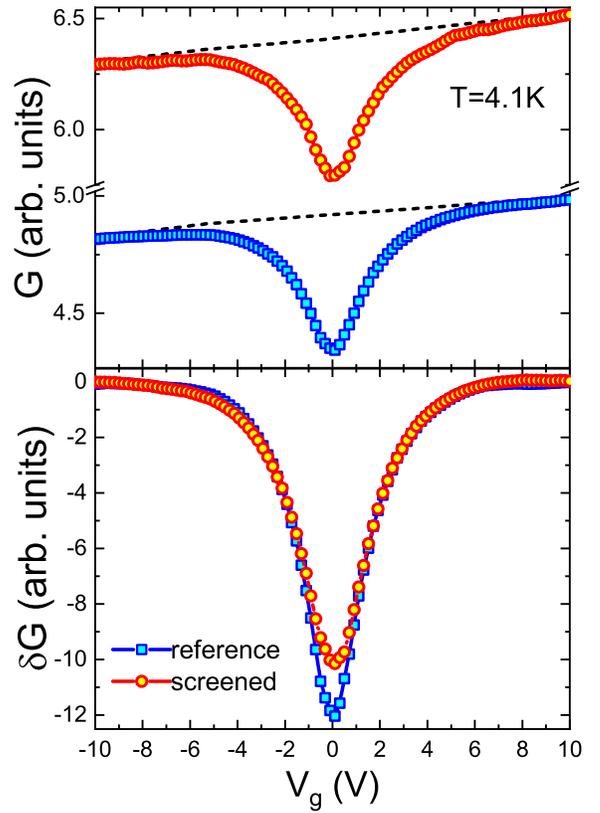}%
\caption{Field-effect plots comparing G(V$_{\text{g}}$) for a reference
(squares) vs. screened (circles) In$_{\text{x}}$O samples from the same
evaporation batch as in Fig.2. The bottom plate is an attempt to scale the
MD's by a constant factor showing a reasonable fit for the "wings" but the
screened-MD falls short of the reference-MD for the bottom part of the
G(V$_{\text{g}}$).}%
\end{figure}

The top plate of Fig.2 show the raw data for the field-effect G(V$_{\text{g}}%
$) of these samples. Two features are observed in this figure; an asymmetric
component characterized by $\partial$G(V$_{\text{g}}$)/$\partial$V$_{\text{g}%
}$%
$>$%
0 that reflects the increased thermodynamic density-of-states with energy (the
thermodynamic field-effect), and a cusp-like dip centered at V$_{\text{g}}$=0
where the system was allowed to relax before sweeping the gate voltage (the
memory-dip). By subtracting from each plot the respective thermodynamic
G(V$_{\text{g}}$) component, one gets the two MD's that, after multiplication
by a constant are shown to have the same shape despite the large disparity in
their resistance.

By contrast, the MD's of the reference-screened samples fail to show similar
data collapse. Figure 3 shows the results of an attempt to match the
memory-dips for a specific couple. In this case, factor-scaling the data for
the two memory-dips is possible for most of the range of G(V$_{\text{g}}$) but
not near its equilibrium point where the screened dip falls short of the
reference. Figure 4 shows however that the current-field characteristics of
these samples is nearly identical even deeper into the non-ohmic regime and
there is no sign of a current-short from the active sample to the
screening-layer.%
\begin{figure}[ptb]%
\centering
\includegraphics[
height=4.0248in,
width=3.039in
]%
{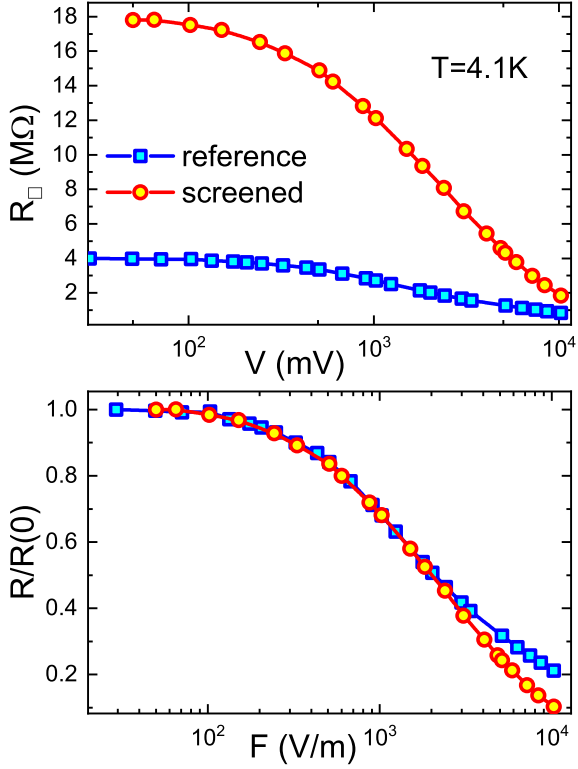}%
\caption{Top plate: The dependence of the sheet-resistance R$_{\square}$ of
the screened (circles) and reference (squares) samples on the applied voltage
(same samples as in Fig.3). Bottom plate: The relative change of these
resistances showing a similar functional dependence.}%
\end{figure}

Note that in this pair, the resistance of the screened sample was larger than
that of the reference. The cutback-shaped MD of the screened sample relative
to the reference has been observed in all six pairs studied in this work, This
was independent of the relative value of the resistances involved. The scaled
results for a pair where the sheet-resistance R$_{\square}$ of the screened
sample is smaller than that of the reference sample are shown in Fig.5 which
depicts the same qualitative features as in Fig.3.
\begin{figure}[ptb]%
\centering
\includegraphics[
height=2.2182in,
width=3.039in
]%
{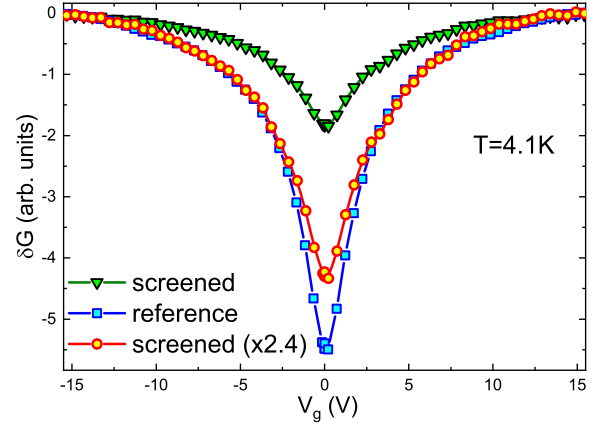}%
\caption{Comparing the MD shape of a reference sample with R$_{\square}$
=6.3M$\Omega$ with a screened sample of the same deposition batch with
R$_{\square}$ =3.1M$\Omega.$}%
\end{figure}
Finally, Fig.6 shows two more reference-screened pairs taken from a single
specific deposition-batch with carrier-concentration \textit{N}$\approx
$1.9x10$^{\text{19}}$cm$^{\text{-3}}$. The figure includes both R$_{\square}%
$(reference)%
$>$%
R$_{\square}$(screened) and R$_{\square}$(reference)%
$<$%
R$_{\square}$(screened) cases as well as an extended range of the field-effect
vs. a higher resolution view of the memory-dip main features.
\begin{figure}[ptb]%
\centering
\includegraphics[
height=4.1208in,
width=3.039in
]%
{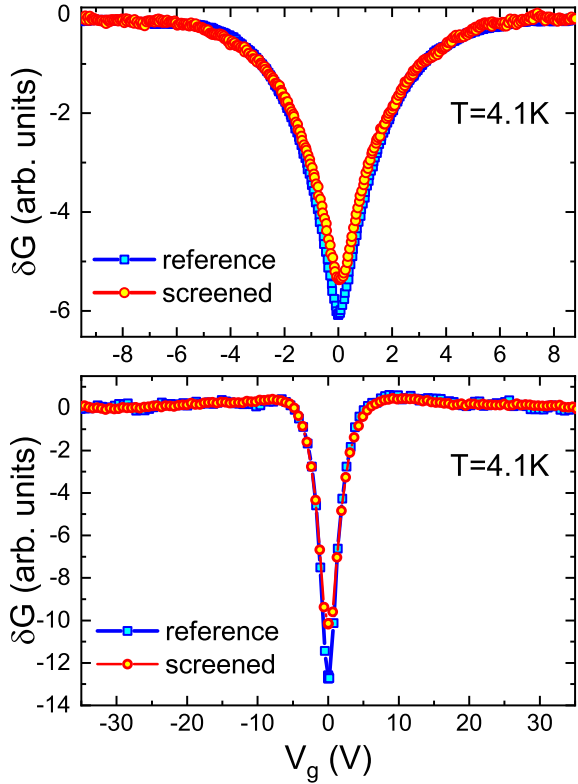}%
\caption{Attempting to scale the functional dependence of the MD of a
reference-screened pair of samples from the same deposition batch
(with\textit{N}$\approx$1.9x10$^{\text{19}}$cm$^{\text{-3}}$). Top plate:
reference sample with R$_{\square}$ =4M$\Omega$, and screened sample with
R$_{\square}$ =40M$\Omega$. Bottom plate: reference sample with R$_{\square}$
=6.7M$\Omega$; screened sample with R$_{\square}$ =45M$\Omega.$}%
\end{figure}
Screening by a nearby metallic plane has been shown to change the spatial
dependence of the Coulomb interaction even in diffusive systems; Some features
of the single-particle DOS found in tunneling experiments on two and
three-dimensional indium-oxide samples differed in both magnitude and
functional dependence from those predicted by simple models of interaction and
disorder \cite{20}. These differences were accounted for by Altshuler et al
based on the image-charges created due to proximity of the sample to the
tunneling electrode \cite{21}. The lack of screening in the
Anderson-insulating phase makes the system more susceptible to the influence
of the nearby metallic-plane. A modified form of the Coulomb interaction is
therefore an expected effect \cite{22}. Indeed, the reduced relative magnitude
of the screened-MD observed in the current experiments is consistent with the
effect of a screening layer on the Coulomb-gap of a 2D system. This effect was
estimated theoretically by Hadley et al \cite{18}. In our six samples the
reduced magnitude of the MD of the screened sample ranged between $\approx
$12\% to $\approx$23\%, which according to \cite{18} are associated with
d\textit{N}$^{\text{1/3}}\approx$2.8 to $\approx$1.6 respectively. For the
carrier-concentrations used in this work \textit{N}=8.7x10$^{\text{19}}%
$cm$^{\text{-3}}$ to 1.9x10$^{\text{19}}$cm$^{\text{-3}}$, these values give d
in the range 8-15nm which in good agreement with the thickness of the
SiO$_{\text{2}}$ spacer used (see section II above). In the six pairs of
screened-reference samples however, it was not possible to see a systematic
dependence on the spacer d. This is probably due to relatively large thickness
variations in these thin films; both SiO2 and In$_{\text{x}}$O have been
tested by AFM which showed thickness fluctuations of the order of $\pm$8\%
\cite{23}.

It is natural to ask how limiting the interaction-range affects how the system
thermalizes after being taken out of equilibrium, attempts to answer this
question are discussed next.

\subsection{Thermalization dynamics}

\subsubsection{Experimental definition of the thermalization-time}

It is rarely possible to ascertain experimentally that a system under
observation is thermalized. One may however monitor the process of the
approach towards equilibrium by following a specific measurable and associate
the state of thermalization with the time where this measured quantity reached
a time-independent value relative to which the system just fluctuates.
Thermalization and relaxation will be used here interchangeably although
technically the time-independent regime may only signal pre-thermalization.

An effective and way to take the system far from equilibrium and observe the
ensuing relaxation is the `gate-protocol'. In this protocol a nonequilibrium
state is created by switching the gate-voltage V$_{\text{g}}$ from an
equilibrium value V$_{\text{eq}}$ to a new one V$_{\text{n}}$. This process is
reflected in the appearance of excess-conductance $\Delta$G(t) that decays
slowly with time. An example for the results obtained with this protocol is
shown in Fig.7.%
\begin{figure}[ptb]%
\centering
\includegraphics[
height=2.1499in,
width=3.039in
]%
{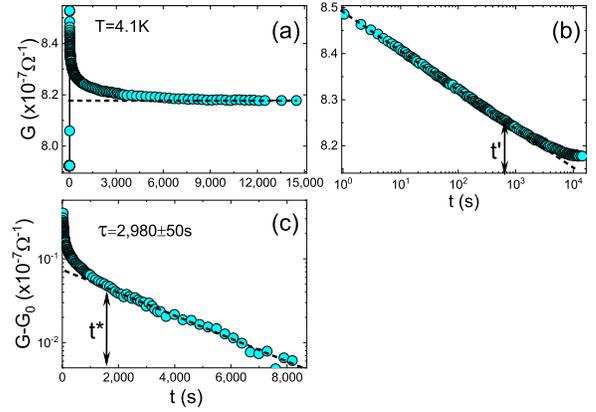}%
\caption{Results of using the gate-protocol (see text) on a sample with
R$_{\square}$=1.2M$\Omega.$ (\textbf{a}) Conductance as function of time;
after $\approx$30 seconds of monitoring G under V$_{\text{g}}$=0V the
gate-voltage was swept to V$_{\text{g}}$=40V at a rate of 15V/s. The dashed
line is G$_{\text{0}}$, the asymptotic value of the conductance. G$_{\text{0}%
}$=G(40V) differs from G(V$_{\text{eq}}$) due to the component of the
thermodynamic field-effect. (\textbf{b}) Conductance relaxation starting from
the time V$_{\text{g}}$=40V was established showing the extent of the log(t)
dependence (delineated by the dashed line). t' marks the point where G(t)
deviates from the logarithmic dependence. (\textbf{c}) The plot of
G(t)-G$_{\text{0}}$ demonstrating an exponential relaxation law: $\Delta
$G(t)$\propto$exp[-(t/$\tau$)] (dashed line is best fit yielding the
relaxation time $\tau$ for the sample). t* marks the time below which G(t)
deviates from exponential-relaxation.}%
\end{figure}

The relaxation to the equilibrium under the newly established V$_{\text{n}}$
was monitored through the measured $\Delta$G(t). As observed in Fig.7,
$\Delta$G(t)$\propto$-log(t) for several hundred seconds (up to t' in Fig.7b),
and after a time marked as t* (Fig.7c) the relaxation-law reverts to $\Delta
$G(t)$\propto$exp[-t/$\tau$], which defines $\tau$ that will be used here as
the characteristic thermalization-time. As will be shown in the next
paragraph, when properly implemented, this `gate-protocol' is equivalent to
quench-cooling the system from high temperatures. The latter, has the
advantages of being history-free but the thermal-cycle runs the risk of
changing the structure of the sample (and possibly damage it more seriously),
and it also sacrifices the short-time relaxation because one must wait for the
sample and its surroundings (sample stage, thermometer, etc.) to cool to the
bath temperature. The gate-protocol, by contrast, may be safely repeated many
times on the same sample and no `parasitic' heating is involved in the process.

An important caveat when using the gate-protocol is to let the sample reach
equilibrium before changing the gate voltage to a new value to avoid history
dependence \cite{24}. As a check on this point, we compared the relaxation
time of a sample by both, a thermal-quench and the gate-protocol. The results,
shown in Fig.8, demonstrate that the relaxation-time $\tau$ based on the
gate-protocol is essentially identical with that based on quench-cooling the
sample. We believe that the gate-protocol can be relied upon to yield the
correct relaxation-time provided the equilibration-time is longer than $\tau$.
For the series of measurements reported below, the samples were equilibrated
under V$_{\text{eq}}$ for at least 12 hours under V$_{\text{eq}}$.%
\begin{figure}[ptb]%
\centering
\includegraphics[
height=2.041in,
width=3.039in
]%
{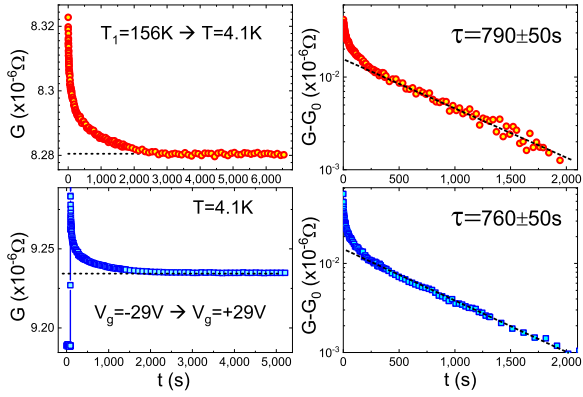}%
\caption{The relaxation dynamics of a In$_{\text{x}}$O film with R$_{\square}$
=0.13M$\Omega$ at 4.1K and \textit{N}$\approx$1.9x10$^{\text{19}}%
$cm$^{\text{-3}}$ tested by two protocols: Top two plates: Using a quench-cool
protocol. Lower two plates: Using the gate protocol. Dashed lines delineate
the equilibrium conductance G$_{\text{0}}$ for each protocol. The relaxation
time $\tau$ is obtained from fit to $\Delta$G(t)$\propto$exp[-(t/$\tau$)] for
the data in the two right-hand-side plots.}%
\end{figure}

\subsubsection{Dynamics of screened-reference samples}

Comparing between screened and the reference samples is more problematic when
it comes to dynamics than the difficulties mentioned above with regard to the
effect on the shape of the memory-dip. The latter is independent of the
disorder; the MD shape is the same even when the sample R$_{\square}$ changes
by order of magnitude (see Fig.2 above) while the dynamics is quite sensitive
to the sample disorder \cite{17} as will be demonstrated below. Figure 9 shows
$\Delta$G(t) for the asymptotic relaxation regime generated by using the
gate-protocol. These data were taken on the same samples used for comparing
the MD shapes in Fig.5 above that, in terms of their R$_{\square}$, is our
best-matched screened-reference pair. The data in Fig.9 clearly suggest that
the relaxation-time of the screened sample is essentially the same as the
reference. Therefore, limiting the range of the Coulomb interaction to
$\approx$8nm does not have a significant effect on the system relaxation-time.
Moreover, $\tau$ of the order of few thousands of seconds is manifestly
possible even \textit{without long-range interaction}. This is a useful piece
of information that should make it easier for theory to finally address the
long-standing question of the slow relaxation times of some electron-glasses
\cite{24}. We return to this issue after discussing the results of the
dynamics as function of disorder.%
\begin{figure}[ptb]%
\centering
\includegraphics[
height=2.1491in,
width=3.0381in
]%
{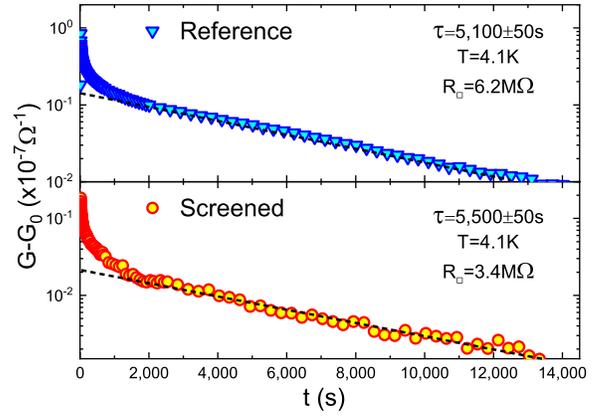}%
\caption{The asymptotic relaxation curves for a specific In$_{\text{x}}$O
screened-reference pair (\textit{N}$\approx$1.9x10$^{\text{19}}$%
cm$^{\text{-3}}$) using the gate protocol under identical conditions. Dashed
lines are best fits to $\Delta$G(t)$\propto$exp[-(t/$\tau$)] that yielded the
respective relaxation time shown in the figures.}%
\end{figure}

\subsubsection{Disorder vs. Interaction}

To get a better picture of the dynamics we expanded a preliminary study of
In$_{\text{x}}$O films with low carrier-concentration \textit{N}%
=8.7x10$^{\text{19}}$cm$^{\text{-3}}$ \cite{25} by measuring 13 samples from
the batch with \textit{N}=1.9x10$^{\text{19}}$cm$^{\text{-3}}$ with which most
of the screen-reference samples studied here were made.%

\begin{figure}[ptb]%
\centering
\includegraphics[
height=2.1741in,
width=3.039in
]%
{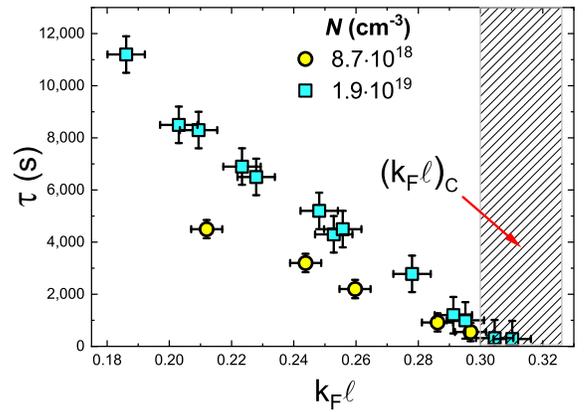}%
\caption{The dependence of the relaxation times $\tau$ (as defined in Figs.7
and 8 above) on the disorder parameter k$_{\text{F}}\ell$ near the
critical-regime of the metal-to-insulator transition (marked by the hatched
area).}%
\end{figure}
Figure 10 shows the relaxation time $\tau$ (defined by the
exponential-relaxation regime of the gate-protocol) as function of the
dimensionless parameter k$_{\text{F}}\ell$. As in other studies \cite{26,27},
k$_{\text{F}}\ell$=(3$\pi^{\text{2}}$)$^{\text{2/3}}\text{\textperiodcentered
}\hbar$\textperiodcentered$\sigma_{\text{RT}}\text{\textperiodcentered
}e^{\text{-2}}$\textperiodcentered\textit{N}$^{\text{-1/3}}$ was taken as the
measure of the quenched disorder ($\sigma_{\text{RT}}$ is the sample
conductivity at room-temperature).

There are two interesting features that emerge from the data; First, the
relaxation-time decreases with k$_{\text{F}}\ell$ and tends to zero roughly at
the disorder range where the system undergoes the metal to insulator
transition. The critical value of disorder (k$_{\text{F}}\ell$)$_{\text{C}}$
for the metal-insulator transition was independently measured for two versions
of the material with \textit{N}$\approx$10$^{\text{21}}$cm$^{\text{-3}}$
\cite{26} and \textit{N}$\approx$10$^{\text{19}}$cm$^{\text{-3 }}$\cite{27}
yielding in both (k$_{\text{F}}\ell$)$_{\text{C}}$=0.31$\pm$0.03. \newline
That the glassy features end at the transition is an important finding; it
supports the conjecture that the slow relaxation is an electronic effect
rather than reflecting structural defects. Note that the reduction of $\tau$
with k$_{\text{F}}\ell$ is achieved in In$_{\text{x}}$O by thermal-annealing.
Changes in the structural properties of the material during the annealing
process were extensively studied in \cite{17} by electron-diffraction,
energy-dispersive spectroscopy, x-ray interferometry, and optical techniques.
The study revealed that the change in the resistance from the as-deposited
deeply-insulating state all the way to the metallic regime is mainly due to
increase of the material \textit{density}. In particular, the samples retained
their amorphous structure and composition throughout the entire process.
Moreover, the dynamics associated with structural changes monitored during
annealing and recovery of the samples was qualitatively different than that of
the electron-glass and did not change its character throughout the entire
range of disorder. The diminishment of $\tau$ with k$_{\text{F}}\ell$ cannot
then be identified with the elimination of some peculiar structural defects.
\newline Secondly, the relaxation-time is not a function of just k$_{\text{F}%
}\ell$; it appears that it also depends on the carrier-concentration
\textit{N}. Indeed, the exponential relaxation regime (which allows an
unambiguous definition of $\tau$) become quickly out of reach for samples when
\textit{N}$\geq$5x10$^{\text{19}}$cm$^{\text{-3}}$. The data in Fig.10 seem to
suggest a scaling relation of the form:%
\[
\tau=\tau\text{(\textit{N})}\cdot\text{[k}_{\text{F}}\ell\text{-(k}_{\text{F}%
}\ell\text{)}_{\text{C}}\text{];}~\text{for k}_{\text{F}}\ell\geq
\text{(k}_{\text{F}}\ell\text{)}_{\text{C}}%
\]
where the prefactor $\tau$(\textit{N}) presumably increases with
carrier-concentration. One may surmise that the dependence on \textit{N} may
be the effect of interactions. The logic is based on the realization that, due
to lack of electronic screening of the Anderson-insulator, higher density of
carriers enhances the strength of interaction. While it is plausible that
interactions in the localized system get stronger with \textit{N}, it is not
necessarily the main (or the only) reason for slower relaxation \cite{17}. It
is here that the issue of separating effects of Coulomb interaction from of
the effect of disorder that presents a frustrating problem because interaction
and disorder both increase with \textit{N}. Actually, a viable cause for
$\tau$ increasing with \textit{N} is the higher degree of disorder in samples
that have higher carrier-concentrations. Note that a pre-condition for the
electronic system to exhibit slow relaxation is Anderson-localization
\cite{17}. This requires that the disorder energy $\mathcal{W}$ has to be
larger than the Fermi energy E$_{\text{F}}$ by a certain factor \cite{28,29}.
All other things being equal, a system with larger carrier-concentration
\textit{N} must be more disordered to be Anderson-localized and thus has
larger $\mathcal{W}$. This, in turn, will \textit{exponentially }slowdown the
inter-site transitions, whether activated or through tunneling.

The way that Coulomb interactions affect thermalization dynamics is less
clear. Interactions may modify transition rates through reduction of the
density of states and many-particle transitions may be involved in the process
but it is hard to find experimental evidence that may be uniquely related to
these mechanisms. A large magnitude of memory-dip, suggestive of a more
dominant role of interactions, is actually found in low-\textit{N} systems
where dynamics is relatively fast as found in the present study. This however
does not mean that interactions act to speed-up thermalization, rather it
shows that the disorder effect (being weaker in low-\textit{N}) is more
important. A possible example for enhanced carrier-concentration without the
accompanying increase of disorder was observed in GeSbTe samples in their
persistent photoconductive state \cite{30}. This caused an enhanced magnitude
of the memory-dip, which was interpreted as an interaction effect \cite{30}.
It also slowed-down the dynamics but interaction is not the only possible
mechanism for it; the slow decay of the photo-induced carriers may be the more
mundane reason.%

\begin{figure}[ptb]%
\centering
\includegraphics[
height=2.1724in,
width=3.039in
]%
{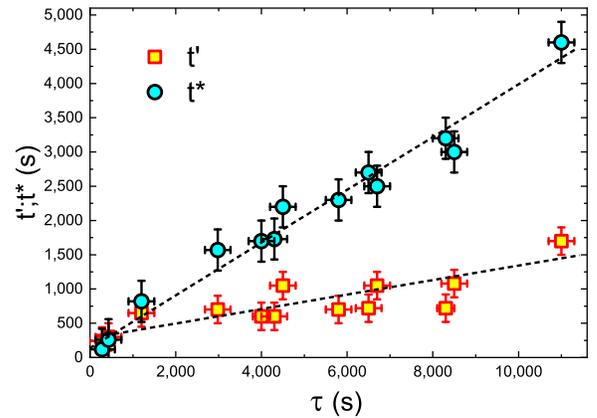}%
\caption{The end-times for the logarithmic and simple exponential relaxation
t' and t* respectively as a function of $\tau$ for the samples studied by the
gate-protocol in Fig.10.}%
\end{figure}

The transition to the exponential relaxation at long times, as observed for
example in Fig.7 above, is expected. A log(t) relaxation is limited to
intermediate times; it has to cross over to a different form for both short
and long times \cite{31}. The transition from a logarithmic to the
exponential-relaxation regime is preceded by a more complicated time
dependence which perhaps resembles the fast relaxation observed in
phosphorous-doped silicon \cite{32}. This region extends over a time-period
that grows monotonically with disorder \cite{33} as is shown in Fig.11.

The dependence of the dynamics on k$_{\text{F}}\ell$ may be summarized as
follows: As k$_{\text{F}}\ell$ increases and the system approaches the
diffusive regime, the rate-distribution, that controls the relaxation from an
excited state, gets narrower due to reduction of the lowest transition-rates
(associated with the $\tau$ deduced from the exponential-relaxation regime).
Concomitantly, the range over which logarithmic relaxation is observed shrinks
linearly with k$_{\text{F}}\ell$ (Fig.11).

It is intriguing that the time-period for the log(t) relaxation in
electron-glass may extend over almost six decades \cite{24} without a sign of
a crossover. To account for such an extensive range one has to assume a fairly
uniform distribution of transition-rates over a wide frequency range. It seems
obvious that a main ingredient in the underlying mechanism is sufficiently
strong disorder, but it probably also involves many-body effects \cite{34}.
The current study demonstrated that electronic relaxation extending over
thousands of seconds is a viable possibility without the long-range part of
the Coulomb interaction playing a significant part (and therefore the DOS at
the Fermi-energy must be finite even at T=0).

It is harder to assess the contribution of short and medium-range Coulomb
interaction to the dynamics. One might argue that the faster dynamics observed
as the system approaches the metallic regime may, at least in part, be due to
the enhanced dielectric-constant that in turn weakens the interaction. The
dielectric-constant in the localized state is expected to increase
significantly near the transition \cite{35}. However, the functional
dependence of $\tau$(k$_{\text{F}}\ell$) shown in Fig.10 does not exhibit a
change from the linear dependence as the transition is approached. Therefore,
this scenario is not supported by our experiments. Interactions are more
likely to play a significant role in the ultra-slow processes that are
necessary to reach the true ground-state of the system, a process that
presumably hinges on many-particle transitions \cite{36}.

There are other mechanism that may contribute to stretch the transition-rates
distribution and afford an extended log(t) dependence. Reduction of
transition-rates relative to the "bare" rates controlled by disorder may occur
for non-local interactions. These may bring into play additional constraints
as well as effects related to coupling of the tunneling charge to other
degrees of freedom (polaronic-effects, and the orthogonality-catastrophe
\cite{37,38,39}). Resolution of these issues remain a challenge to theory.

\begin{acknowledgments}
Illuminating discussions with A. Vaknin and M. Schechter are gratefully
acknowledged. This research has been supported by a grant No 1030/16
administered by the Israel Academy for Sciences and Humanities.
\end{acknowledgments}


\begin{thebibliography}{99}
\bibitem {33}This part of the relaxation may be fitted to a
stretched-exponential function, $\Delta$G(t)=A$\cdot$exp[-(t/$\tau^{\prime}%
$)$^{\beta}$] by adjusting the parameters $\tau^{\prime}$and $\beta$. For
example, the data for the intermediate range of 15s$\leq$t$\leq$%
5x10$^{\text{3}}$s in Fig.7c may be fitted with the three parameters
A=3.3x10$^{\text{-8}}\Omega^{\text{-1}}$, $\tau^{\prime}$=185s, and $\beta$=0.33.
\end{thebibliography}
\end{document}